\documentclass[prb,twocolumn,superscriptaddress,floats,showpacs]{revtex4}

\usepackage[dvips]{graphicx}
\newcommand{\Sn}{\mbox{$^{119}$Sn}}
\newcommand{\Fe}{\mbox{$^{57}$Fe}}

\begin{document}

\title
{Measuring velocity of sound with nuclear resonant inelastic x-ray scattering}

\author{Michael Y. Hu}
\email{myhu@anl.gov}
\affiliation
{HP-CAT and Carnegie Institution of Washington, 
Advanced Photon Source, Argonne, IL 60439}

\author{Wolfgang Sturhahn}
\author{Thomas S. Toellner}
\affiliation
{Advanced Photon Source, Argonne National Laboratory, Argonne, IL 60439}

\author{Philip D. Mannheim}
\affiliation
{Department of Physics, University of Connecticut, Storrs, CT 06269}

\author{Dennis E. Brown}
\affiliation
{Department of Physics, Northern Illinois University, DeKalb, IL 60115}

\author{Jiyong Zhao}
\author{E. Ercan Alp}
\affiliation
{Advanced Photon Source, Argonne National Laboratory, Argonne, IL 60439}

\date{Jan. 17, 2003}

\begin{abstract}

Nuclear resonant inelastic x-ray scattering is used to measure 
the projected partial phonon density of states of materials.
A relationship is derived between the low-energy part of this
frequency distribution function and the sound velocity of materials.
Our derivation is valid for harmonic solids with Debye-like
low-frequency dynamics.
This method of sound velocity determination is applied to
elemental, composite, and impurity samples which are representative
of a wide variety of both crystalline and noncrystalline materials.
Advantages and limitations of this method are elucidated.

\end{abstract}



\maketitle

Mechanical properties form an important part of our understanding 
of condensed matter.
In many areas of science, measurements of sound velocity are used
to study materials of both natural occurence and artificial fabrications.
For example, in the field of geophysics, the sound velocity is the most
direct information we have about the Earth's interior.
The standard approach to learn about the composition and structure of
the Earth's interior entails measurements of sound velocities of
candidate compounds.
The results are then compared to seismological data to exclude or
confirm a particular compound.
In the following, we will describe the use of nuclear resonant inelastic
x-ray scattering (NRIXS) to measure the velocity of sound.

The NRIXS method was introduced to probe the lattice dynamics of
materials by employing low-energy nuclear resonances.~\cite{SYK+95,STA+95}
In NRIXS experiments, only signals from nuclear resonance absorption
are monitored, and for this reason the extracted quantity is specific
to the resonant isotope.
This technique provides the phonon excitation spectrum as seen by the
probe nuclei,~\cite{SK99a,CS99a,HST+99} and in most cases one can extract
the partial vibrational frequency distribution, a function often
referred to as the partial phonon density of states (PDOS).
The NRIXS method has been applied to various materials, e.g.,
thin films and multilayers,~\cite{RST+99,KS99,RKS+2001}
nanoparticles,~\cite{BAA+97,PBC+2002}
crystals with impurities,~\cite{SKK+2000}
organic molecules,~\cite{PWT+99,PBH+2001,RDP+2002a,RDP+2002b}
proteins,~\cite{SDS+2001,AKO+2002}
samples under high pressures,~\cite{SMH+2001}
and samples of geophysical interests.~\cite{LGC+2000,MXS+2001}
Most of these samples are compounds, and, while the obtained PDOS
gives only part of the lattice dynamics, the low-energy portion of
the PDOS provides the Debye sound velocity of the whole sample.
We will now show that, due to universal features of acoustic modes of
harmonic solids, the low-energy portion of the PDOS is related to
the Debye sound velocity in a simple way.

The normalized phonon density of states is defined by
\begin{equation}
 \mathcal{\nu}(E) = 
 \frac{1}{3N}
 \sum_{l}^{3N} \delta\Big(E - E_l \Big) \:\:,
\label{Eq:DOS}
\end{equation} 
where the energy eigenstates of lattice vibrations $E_l$ are labeled
by quantum number $l$, and $N$ is the total number of atoms in the solid.
In the harmonic lattice approximation, a PDOS, which is more relevant to
NRIXS experiments, is given by~\cite{SK99a}
\begin{equation}
 \mathcal{D}(E,\hat{\mathbf k}) = 
 \frac{1}{\tilde{N}} \sum_{\nu=1}^{\tilde{N}}  \,
 \frac{1}{N} \sum_{l=1}^{3N} \,
 | \hat{\mathbf k} \cdot {\mathbf e}^{\nu}_{\,l} |^2 \,
 \delta (E - E_l) \:\:,
\label{Eq:pDOS}
\end{equation} 
where $\nu$ enumerates resonant nuclei, $\tilde{N}$ is the total number
of resonant nuclei, $\hat{\mathbf k}$ is a unit vector in the incident
photon direction, and ${\mathbf e}^{\nu}_{\,l}$ are phonon polarization
vectors.
Equation (\ref{Eq:pDOS}) shows that the vibrational polarizations are
projected onto the incident photon direction and in particular
the vibrational modes with polarization perpendicular to the direction of
the incident photon do not contribute.
In the case of a single crystal, the measured vibrational properties
become dependent on the incident photon direction and were called
``projected,''~\cite{CRB+97,KCR98}
whereas in cases of polycrystalline or isotropic samples,
the measured spectrum is an average over all directions.
For a crystal in which resonant nuclei occupy only equivalent lattice sites,
the quantity that can be extracted from NRIXS experiments is exactly
described by Eq.~(\ref{Eq:pDOS}).
When the resonant nuclei occupy different sites, what can be extracted is 
an approximation of Eq.~(\ref{Eq:pDOS}).
The approximation is based on an average of phonon spectra
for these different lattice sites.
The closure conditions of the phonon polarization vectors
guarantee the normalization of $\mathcal{D}(E,\hat{\mathbf k})$, i.e.,
its integration over all phonon energies is one.
The orthonormality and closure conditions are given by
\begin{eqnarray}
 \frac{1}{N} \, \sum_{\mu=1}^{N}  \sum_{\alpha=1}^{3} \; 
 \left(e^{\mu\alpha}_{\,l}\right)^* e^{\mu\alpha}_{\,l'}
  & = & \delta_{ll'} \, ,  \label{Eq:otho} \\
 \frac{1}{N} \, \sum_{l=1}^{3N} \, 
 \left(e^{\mu\alpha}_{\,l}\right)^* e^{\nu\beta}_{\,l}
  & = & \delta_{\mu\nu} \, \delta_{\alpha\beta} \, ,
\end{eqnarray} 
where $\alpha$, $\beta$ denote the spatial components.
These conditions hold for any harmonic solid, and the polarization vectors have
to be specified for every atom in the solid, in contrast to the crystal case, 
where they can be reduced to a much smaller set for atoms 
in one unit cell only.
The deviation of $D(E,\hat{\mathbf k})$ from $\nu(E)$ 
is contained in the behavior of the
phonon polarization vectors ${\mathbf e}^{\mu}_{\,l}$ and
can be expressed in terms of a modulating function,
\begin{equation}
  \mathcal{D}(E,\hat{\mathbf k}) 
  =  \chi(E,\hat{\mathbf k}) \, \mathcal{\nu}(E) \, ,
\label{Eq:mod}
\end{equation} 
which we now determine for low-energy vibration modes.

In a crystal, the acoustic modes form three branches described by
phonon momentum.
For disordered solids, we still expect hydrodynamic modes on length scales
that are large compared to length scales characterized by inhomogeneities
in the material.
In the appendix, we discuss the properties of such hydrodynamic modes.
Our results show that linearly dispersing plane-wave modes exist on long
length scales, and that these modes are described by momentum $\mathbf{q}$
and branch number $s$.
Energies and atomic displacements associated with these modes
take the following form
\begin{eqnarray}
E_{\mathbf{q}s}&=&c_{\mathbf{\hat{q}}s}\,\hbar q \nonumber \\
\mathbf{u}_{\mathbf{q}s}^\mu&=&
\alpha_{\mathbf{q}s}\,\mathbf{\hat{p}}_{\mathbf{\hat{q}}s}\,
e^{i\mathbf{q\cdot r}_{\mu}}\:\:,
\label{Eq:disp}
\end{eqnarray} 
where $c_{\mathbf{\hat{q}}s}$ is the sound velocity,
$\mathbf{\hat{p}}_{\mathbf{\hat{q}}s}$ is a normalized polarization
vector, and $\mathbf{r}_{\mu}$ are the atomic positions.
The normalization factor $\alpha_{\mathbf{q}s}$ can be
determined as follows.
For a normal mode $l$ with energy $E_l$, the atomic displacements
are given by~\cite{SK99a}
\begin{equation}
\mathbf{u}^{\mu}_l =
\frac{\hbar}{\sqrt{2\,m_\mu E_l}}\:\mathbf{e}^\mu_l\:\:,
\end{equation}
where $\mathbf{e}^\mu_l$ are phonon polarization vectors
introduced in Eq.\,(\ref{Eq:pDOS}) and $m_\mu$ is the mass
of atom $\mu$.
A comparison with Eq.\,(\ref{Eq:disp}) under consideration
of the normalization condition Eq.\,(\ref{Eq:otho}) results in
$\alpha_{\mathbf{q}s}=\hbar/\sqrt{2\,m\,E_{\mathbf{q}s}}$,
where $m$ is the average atomic mass.
For the low-energy, hydrodynamic modes, we can therefore write
\begin{equation}
{\mathbf e}^{\mu}_{l} = {\mathbf e}^{\mu}_{\mathbf{q}s} =
\sqrt{\frac{m_\mu}{m}}\:\mathbf{\hat{p}}_{\mathbf{\hat{q}}s}\,
e^{i\mathbf{q\cdot r}_{\mu}}\:\:.
\label{Eq:pol}
\end{equation}

In the small $q$, low-energy regime, we rewrite Eq.~(\ref{Eq:pDOS}) by
replacing the summation over phonon modes with an integration,
$\sum_{l} \, \rightarrow \, V/(2\pi)^3 \sum_{s} \int q^2dq d\Omega_q$,
and we substitute Eqs.~(\ref{Eq:disp}) and~(\ref{Eq:pol}) 
into Eq.~(\ref{Eq:pDOS}) to obtain
\begin{equation}
\chi(\hat{\mathbf k}) = 
\frac{\tilde{m}}{m}\,v_D^3 \sum_{s=1}^3
 \int \frac{d\Omega_q}{4\pi} \,
   \frac{(\mathbf{\hat{k}\cdot\hat{p}}_{\mathbf{\hat{q}}s})^2}
        {c_{\mathbf{\hat{q}} s}^3}\:\:,
\label{Eq:pDOSle}
\end{equation} 
where $\tilde{m}$ is the mass of the nuclear resonant isotope, and
the Debye velocity $v_D$ is defined as an average over all sound
velocities
\begin{equation}
\frac{1}{ v_D^3}= \frac{1}{3}
 \sum_{s=1}^3 \int \frac{d\Omega_q}{4\pi} \,
\frac{1}{c_{\mathbf{\hat{q}} s}^3}\:\:.
\label{Eq:debyevel}
\end{equation} 
We see that for small energies, the modulating function becomes energy independent.
If we define a projected sound velocity similarly by
\begin{equation}
\frac{1}{ v_\mathbf{\hat{k}}^3} =  \sum_{s=1}^3 
\int \frac{d\Omega_q}{4\pi} \,
\frac{(\mathbf{\hat{k}\cdot\hat{p}}_{\mathbf{\hat{q}}s})^2}
{c_{\mathbf{\hat{q}} s}^3}\:\:,
\label{Eq:paravel}
\end{equation}
we obtain the simple relationship
\begin{equation}
\chi(\mathbf{\hat{k}}) = 
\frac{\tilde{m}}{m}\,\left(\frac{v_D}{v_\mathbf{\hat{k}}}\right)^3\:\:.
\label{Eq:debyePDOS}
\end{equation}
For an isotropic sample, the sound velocity does not have directional
dependence, and we can further simplify Eq.~(\ref{Eq:debyevel}) and
(\ref{Eq:paravel}) to identify $v_\mathbf{\hat{k}}=v_D$.
In the case of a polycrystalline sample, averaging over all nuclear resonant sites 
in Eq.~(\ref{Eq:pDOS}) is equivalent to averaging Eq.~(\ref{Eq:pDOSle})
over all directions of ${\mathbf k}$.
Thus, in both cases, we have
\begin{equation}
\chi = \frac{\tilde{m}}{m} \, .
\label{Eq:chiave}
\end{equation} 
Finally, for an isotropic or a polycrystalline sample,
in the low-energy regime Eq.~(\ref{Eq:pDOS}) becomes
\begin{equation}
 \mathcal{D}(E) = \,
 \left( \frac{\tilde{m}}{m} \right)
 \frac{E^2}{2 \pi^2 \hbar^3 n v_D^3} \, ,
\label{Eq:debye}
\end{equation}
where $n$ is the density of atoms.
Equation~(\ref{Eq:debye}) has appeared in a similar form in a NRIXS study of 
myoglobin and related biological compounds.~\cite{AKO+2002}
Here we have given a derivation of Eqs.~(\ref{Eq:chiave}) and (\ref{Eq:debye}).
In addition, Eq.~(\ref{Eq:debyePDOS}) shows the dependence
of the modulation function on the photon direction for anisotropic
samples.
A ``mean sound velocity" was 
defined in the context of NRIXS in a previous study.~\cite{KCR98}
It is identical to  $v_\mathbf{\hat{k}}$ for isotopically pure samples,
for which the PDOS becomes total DOS.
Here we have shown that the sound velocity $v_\mathbf{\hat{k}}$ of a sample 
in general is obtained by applying a correction factor,
the cube root of the modulation factor as given in Eq.~(\ref{Eq:chiave}),
to the ``mean sound velocity" of Ref.~\onlinecite{KCR98}.

The PDOS of a variety of samples has been measured by NRIXS.
Here we show three examples representing crystals (bcc iron, hematite)
and materials with point defects (\Sn$_{0.01}$Pd$_{0.99}$).
The crystalline samples are translationally invariant, 
and thus for them the basis of modes is in fact given by momentum
eigenstates with polarization vectors which are known to obey
Eq.~(\ref{Eq:pol}) explicitly.~\cite{MMWI71}
While the point-defect case is expressly not translationally invariant, 
by using the lattice Green's function technique, it is
still possible to find closed form expressions for the
polarization vectors, with Eqs.~(\ref{Eq:chiave})~and~(\ref{Eq:debye})
again being found to follow. 
We shall describe the details of this specific calculation 
elsewhere,~\cite{HMA+2002} while noting here that the modulation to the 
low-energy part of host lattice DOS is found to depend only on the 
mass ratio even in the event of a force constant change at
the defect site.~\cite{Mannheim68}

\begin{figure}
\hspace{0mm}
\includegraphics*[scale=1.0]{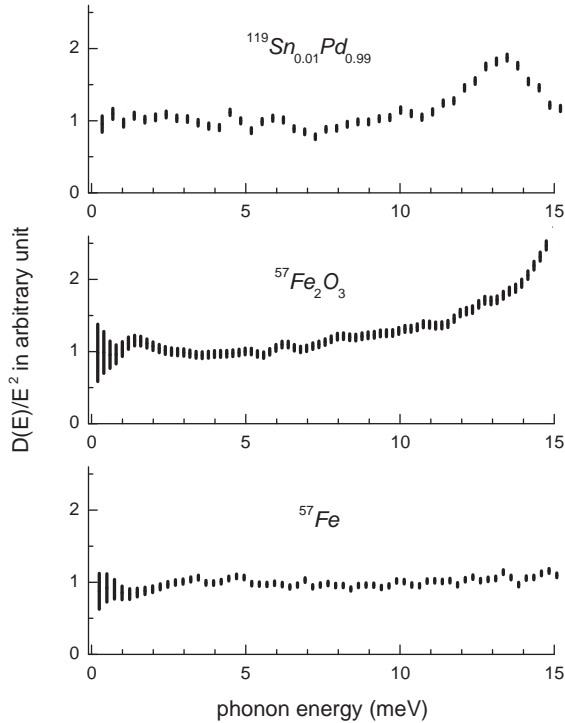}
\caption
{
 The measured PDOS divided by energy squared.
 The size of symbols indicates the statistical error bar derived from signal counts.
 These samples were measured with resolutions of
 0.85\,meV for \Sn$_{0.01}$Pd$_{0.99}$, 0.6\,meV for \Fe$_2$O$_3$, and
 1\,meV for bcc iron. 
 \label{fig1}
}
\end{figure}

The low-energy region of the PDOS divided by energy squared
is displayed for the three samples in Fig.~\ref{fig1}.
Bcc iron 95\%\ enriched in \Fe\ is an example of the limiting case 
where $\tilde{N} \sim N$, and NRIXS provides the total rather than
the partial DOS.
Hematite enriched with \Fe\ represents a situation, in which
the resonant nuclei form only a part of the unit cell.
The  \Sn$_{0.01}$Pd$_{0.99}$ sample approaches another
limiting case in which the nuclear resonant isotopes
occupy only a very small portion of lattice sites.
In this limit, $\tilde{N}\ll N$, the average atomic mass $m$ in the above
equations is well approximated by the mass of a host lattice atom, 
and the sound velocity obtained is that of the pure host, rather than of 
the host/impurity system.
These samples thus represent a very broad range of nuclear resonant
isotope concentrations.

We extracted numerical values for the velocities of sound by
averaging $\mathcal{D}(E)/E^2$ obtained from measured data
in the region from zero to 5\,meV according to Eq.~(\ref{Eq:debye}).
The results are tabulated in Table~\ref{Tbl:vc},
where we also compare our values with sound velocities
from other sources as explained in the footnotes of the table. 
For iron and palladium, the results from NRIXS measurements are
within the range of sound velocities that were obtained by
other means.
In the case of hematite, our results give clearly a lower
value for $v_D$ than is obtained from the measured elastic
constants.
It is difficult for us to judge the reliability of values
for $v_D$ obtained by other methods that usually do not measure
$v_D$ directly but rely on post-experimental data averaging.
Values obtained with the NRIXS method seem to be lower or on
the lower end of ranges given by other authors.
This might be due to the fact that the phonon spectrum
up to 5\,meV (corresponding to a frequency of 1.2\,THz)
is used to derive the sound velocity.
At such high frequencies, the phonon dispersion may already
be nonlinear, which would typically lead to a reduction in the
value of the obtained $v_D$.
Improvements in energy resolution could make smaller phonon energies
accessible, which would potentially provide a better measure of
sound velocity.

\begin{table}[t]
{
\caption{\label{Tbl:vc}Comparison of velocities of sound,
         together with the factor ($\tilde{m}/m$) in 
         Eqs.~(\ref{Eq:chiave})~and~(\ref{Eq:debye}).}
\begin{ruledtabular}
\begin{tabular}{cccc}
         & velocity of sound (m/s) & $\tilde{m}/m$\\  
\hline
Palladium&2193$\pm$35\footnotemark[1] & 1.12\\
         &2104\footnotemark[2] & \\
         &2372\footnotemark[3] & \\
\hline
Hematite &4279$\pm$84\footnotemark[1]  & 1.76\\
         &4653\footnotemark[4] & \\
\hline
Iron &3488$\pm$48\footnotemark[1] & 1.00\\
         &3412\footnotemark[2] & \\
         &3707\footnotemark[3] & \\
\end{tabular}
\end{ruledtabular}
\footnotetext[1]{Our results from NRIXS.}
\footnotetext[2]{The lower limits from ref.~\onlinecite{SW71}.}
\footnotetext[3]{The upper limits from ref.~\onlinecite{SW71}.}
\footnotetext[4]{Calculated with bulk and shear moduli from ref.~\onlinecite{LS68}.}
}
\end{table}

We have relied critically on the assumption of 
Debye behavior at small phonon energies.
In particular, we require Debye behavior to extrapolate to typical sound
frequencies ($10^4$\,Hz) from the THz range, which is accessible 
to NRIXS.
However, in Fig.~\ref{fig1}, we see deviations from Debye behavior,
which would correspond to a horizontal line.
Besides nonlinearities in the phonon dispersion as mentioned above,
the removal of the elastic contributions to the NRIXS spectra can
be a source of systematic uncertainties.~\cite{HST+99,Sturhahn00,KC2000}
Also the NRIXS technique relies on harmonic behavior of the sample
to extract the PDOS from the measured data.~\cite{Sturhahn00,KC2000}
All these possibilities may contribute to the deviations seen in
Fig.~\ref{fig1}.
In future studies aimed at low-frequency dynamics, we suggest
measuring the resolution function simultaneously by nuclear forward
scattering to improve the reliability of peak substraction.
In any case, a highly accurate resolution function should be
available.
Besides the peak-subtraction procedure and the energy resolution
(usually defined as FWHM), the shape of the resolution function
is also of major importance.
The access to small excitation energies is greatly improved if the
tails of the resolution function can be minimized by design of the
x-ray monochromator.~\cite{Toellner00}
Furthermore, difficulties in the peak removal usually become more
severe with increasing ratio of elastic to inelastic scattering
intensities.
In many cases, the relative strength of the elastic scattering is reduced
by saturation effects in the sample.~\cite{STA+95}
This highly desirable effect is less pronounced for samples with
low concentrations of the resonant isotope or in cases of generally
weak inelastic scattering, e.g., at very low temperatures.
In addition to the special requirements for data collection near
the elastic peak, an accurate measurement of the entire spectrum
is equally important to achieve an accurate normalization.

We have shown that nuclear resonant isotopes can be used to measure
the sound velocity of a solid in the context of the harmonic
approximation and Debye-like low-frequency dynamics.
Our approach is valid even for very low concentrations of the
nuclear resonant isotope, and therefore the probing nuclei provide
information about the host lattice.
We expect that the presented method will have significant impact
in the scientific area of high-pressure research and, in particular,
in the field of geophysics, where sound velocity is of great interest.
Also the strategic placement of resonant nuclei in artificial
structures may provide insight into local atomic motion, which at low
energies is thought to influence the electronic noise in
nanostructure devices.
Thus NRIXS opens another venue to measure sound velocities of solids
and can complement other techniques or even supercede established
methods in those cases where they become too demanding or even impossible.


This work and use of the Advanced Photon Source are supported 
by the U.S. Department of Energy, Basic Energy Sciences, Office of Science, 
under contract No. W-31-109-ENG-38,
and by the State of Illinois under HECA.

\section*{Appendix}
Our study of long-wavelength vibrational excitations starts
with the continuum version of the equations of motion of a
set of harmonically bound atoms, which reads
\begin{equation}
\label{eq:discrete}
\omega^2 m^\mu \mathbf{u}^\mu =
-\sum_{\nu} \Phi^{\mu\nu}\,\mathbf{u}^\nu\:\:,
\end{equation}
where $\Phi^{\mu \nu}$ are the force constant matrices and
$\mathbf{u}^\mu$ is the displacement vector of atom $\mu$ for a
vibrational mode of energy $\hbar\omega$.
We introduce mass and atomic density functions $\rho (\mathbf{x})$
and $\eta (\mathbf{x})$ by
\begin{eqnarray}
\label{eq:density}
\rho (\mathbf{x})&=&
\sum_\mu m^\mu \delta^3(\mathbf{x}-\mathbf{r}_\mu) \nonumber \\
\eta (\mathbf{x})&=&
\sum_\mu \delta^3(\mathbf{x}-\mathbf{r}_\mu)\:\:,
\end{eqnarray}
where $\mathbf{r}_\mu$ is the position of atom $\mu$.
The continuum version that substitutes Eq.\,(\ref{eq:discrete})
is then given by
\begin{equation}
\label{eq:continuum}
\omega^2 \rho(\mathbf{x})\mathbf{u}(\mathbf{x}) =
-\int \eta(\mathbf{x})\Phi(\mathbf{x}-\mathbf{x'})\eta(\mathbf{x'})
\mathbf{u}(\mathbf{x'})\:d^3x'\:\:,
\end{equation}
where we made the usual assumption that the force constant
matrix depends on coordinate differences only.
The original intentions to study long-wavelength excitations are
served best by introducing Fourier transforms, e.g.,
$\mathbf{\tilde{u}}(\mathbf{q})=FT[\mathbf{u}(\mathbf{x})]$,
and eventually expanding for small values of momentum.
The transformed Eq.\,(\ref{eq:continuum}) reads
\begin{eqnarray}
\label{eq:FTcontinuum}
&\omega^2& \int \tilde{\rho}(\mathbf{q}-\mathbf{q'})
\mathbf{\tilde{u}}(\mathbf{q'})\:d^3q' = -\frac{1}{(2\pi)^3}\:\times
\nonumber\\
&& \int \tilde{\eta}(\mathbf{q}-\mathbf{q'})
\tilde{\Phi}(\mathbf{q'})\tilde{\eta}(\mathbf{q'}-\mathbf{q''})
\mathbf{\tilde{u}}(\mathbf{q''})\:d^3q'd^3q''\:\:.
\end{eqnarray}
The quantities $\tilde{\eta}(\mathbf{q})$ and $\tilde{\rho}(\mathbf{q})$
are closely related to the structure function $S(\mathbf{q})$ that is
typically obtained from x-ray or neutron diffraction experiments.
For crystals, these functions are described by a series of very
sharp peaks at values given by the reciprocal lattice vectors.
Disordered or amorphous materials do not produce these sharp peaks
with the exception of the $\mathbf{q}=0$ maximum which is not related
to spatial order.
One can assume that some of the salient features of
Eq.\,(\ref{eq:FTcontinuum}) are captured by retaining the $\mathbf{q}=0$
maximum only, i.e., we approximate
\begin{eqnarray}
\label{eq:FTdensity}
\tilde{\rho}(\mathbf{q})&=&
(2\pi)^3\,\frac{m}{V}\,\delta^3(\mathbf{q}) \nonumber \\
\tilde{\eta}(\mathbf{q})&=&
(2\pi)^3\,\frac{1}{V}\,\delta^3(\mathbf{q})\:\:,
\end{eqnarray}
where $m$ and $V$ are the average mass and volume per atom.
Equation (\ref{eq:FTcontinuum}) simplifies to
\begin{equation}
\label{eq:Scontinuum}
m\,\omega^2 \mathbf{\tilde{u}}(\mathbf{q})=
-\frac{1}{V}\,\tilde{\Phi}(\mathbf{q})\mathbf{\tilde{u}}(\mathbf{q})\:\:,
\end{equation}
and describes a class of solutions that should be common
for all harmonic solids.
The previous equation permits us to label the modes with the value of
$\mathbf{q}$ and a branch index $s$ originating from the tensor
character of the force-constant matrix, i.e., $\omega=\omega_{\mathbf{q}s}$.
We note that $\tilde{\Phi}(0)=0$ follows from the invariance of the
vibrational energy with respect to a displacement identical for all
atoms.
Assuming that the force-constant matrix falls off sufficiently fast
with distance an expansion of Eq.\,(\ref{eq:Scontinuum}) in powers
of $q=|\mathbf{q}|$ is permissible.
The first-order term in this expansion will also vanish under the
reasonable assumption of inversion symmetry of the force-constant matrix.
For hydrodynamic modes, i.e., $q\rightarrow 0$, we may therefore use
\begin{equation}
\label{eq:expansion}
\tilde{\Phi}(\mathbf{q}) = \frac{q^2}{2}\,
\left(
\frac{\partial^2\tilde{\Phi}}{\partial q^2}
\right)_{\!\!q=0}\:\:,
\end{equation}
and long-wavelength vibrational excitations are described by
a modified Eq.\,(\ref{eq:Scontinuum})
\begin{equation}
\label{eq:EScontinuum}
m\,\omega_{\mathbf{q}s}^2 \mathbf{\tilde{u}}(\mathbf{q}) = q^2\,
\tilde{\Theta}(\mathbf{\hat{q}})\,\mathbf{\tilde{u}}(\mathbf{q})\:\:.
\end{equation}
We obtain linearly dispersing modes with energies
$\hbar\omega_{\mathbf{q}s}=c_{\mathbf{\hat{q}}s}\,\hbar q$ with
sound velocities $c_{\mathbf{\hat{q}}s}$ that depend on the
direction of $\mathbf{q}$.
The matrix $\tilde{\Theta}$ is related to the elastic tensor of
the material (see for example ref.~\onlinecite{AM76})
and is given by
\begin{equation}
\label{eq:elmatrix}
\tilde{\Theta}(\mathbf{\hat{q}}) = 
-\frac{1}{2V}\,
\left( \frac{\partial^2\tilde{\Phi}}{\partial q^2} \right)_{\!\!q=0} =
\frac{1}{2N}\sum_\mu (\mathbf{\hat{q}\cdot r}_\mu)^2 \Phi(\mathbf{r}_\mu)
\:\:.
\end{equation}
$N$ is the number of atoms in the sample.
The eigenvalues of $\tilde{\Theta}$ provide us with the sound
velocities, and the corresponding eigenvectors will not depend on $q$.
The displacement field for a mode $\mathbf{q}s$ is then given by
\linebreak
\begin{equation}
\label{eq:displacement}
\mathbf{u}^\mu_{\mathbf{q}s} =
\alpha_{\mathbf{q}s}\,\mathbf{p}_{\mathbf{\hat{q}}s}\,
e^{i\mathbf{q\cdot r}_\mu}\:\:,
\end{equation}
where $\mathbf{\hat{p}}_{\mathbf{\hat{q}}s}$ is a normalized
eigenvector of $\tilde{\Theta}$ describing the polarization of the mode,
and $\alpha_{\mathbf{q}s}$ is an appropriately chosen normalization factor.
Equation (\ref{eq:displacement}) shows that, even for disordered materials,
hydrodynamic modes are equivalent to plane-wave excitations of the
atomic displacements.

\end{document}